\documentclass[12pt]{article}
\usepackage{pdproc} 

  \def\be{\begin{equation}}
\def\ee{\end{equation}}
\def\bea{\begin{eqnarray}}
\def\eea{\end{eqnarray}}
\def\beq{\begin{equation}}
\def\eeq{\end{equation}}
\def\bq{\begin{quote}}
\def\eq{\end{quote}}
\def\gappeq{\mathrel{\rlap {\raise.5ex\hbox{$>$}} {\lower.5ex\hbox{$\sim$}}}}
\def\lappeq{\mathrel{\rlap{\raise.5ex\hbox{$<$}} {\lower.5ex\hbox{$\sim$}}}}

\def\epm#1#2{\hbox{${\lower1pt\hbox{$\scriptstyle +#1$}}
\atop {\raise1pt\hbox{$\scriptstyle -#2$}}$}}

\def\gsim{\mathrel{\rlap{\lower4pt\hbox{\hskip1pt$\sim$}}
    \raise1pt\hbox{$>$}}}         

\def\frac#1#2{{{#1}\over {#2}}}

\def\GeV{{\rm GeV}}

\def\bq{\bar{q}}
\catcode`@=11 
\def\slash#1{\mathord{\mathpalette\c@ncel#1}}
 \def\c@ncel#1#2{\ooalign{$\hfil#1\mkern1mu/\hfil$\crcr$#1#2$}}
\def\lsim{\mathrel{\mathpalette\@versim<}}
\def\gsim{\mathrel{\mathpalette\@versim>}}
 \def\@versim#1#2{\lower0.2ex\vbox{\baselineskip\z@skip\lineskip\z@skip
       \lineskiplimit\z@\ialign{$\m@th#1\hfil##$\crcr#2\crcr\sim\crcr}}}
\catcode`@=12 

\begin{document}
\pagestyle{empty}
\begin{flushright}
{CERN-TH/2001-144}\\
\end{flushright}
\vspace*{5mm}
\begin{center}
{\bf MODELS OF NEUTRINO MASSES AND MIXINGS} \\
\vspace*{1cm} 
{\bf G. Altarelli} \\
\vspace{0.3cm}
Theoretical Physics Division, CERN \\
CH - 1211 Geneva 23 \\
\vspace*{2cm}  
{\bf ABSTRACT} \\ \end{center}
\vspace*{10mm}
\noindent
We briefly review models of neutrino masses and mixings. In view of the existing
experimental ambiguities many possibilities are still open. After an overview of the main
alternative options we focus on the most constrained class of models based on three widely
split light neutrinos within SUSY Grand Unification.

\vspace*{5cm} 

\begin{center}
{\it Talk given at the 9th International Workshop on\\
Neutrino Telescopes, 
Venice, March 6--9, 2001}
\end{center}

\vspace*{3.5cm}

\begin{flushleft} CERN-TH/2001-144 \\
May 2001
\end{flushleft}
\vfill\eject

\setcounter{page}{1}
\pagestyle{plain}




\renewcommand{\thefootnote}{\alph{footnote}}
  
\title{
 MODELS OF NEUTRINO MASSES AND MIXINGS}

\author{ Guido Altarelli}

\address{ Theory Division, CERN
 \\
 CH-1211 Geneva 23, Switzerland\\
 {\rm E-mail: Guido.Altarelli@cern.ch}}

  \centerline{\footnotesize and}



\abstract{We briefly review models of neutrino masses and mixings. In view of the existing
experimental ambiguities many possibilities are still open. After an overview of the main
alternative options we focus on the most constrained class of models based on three widely
split light neutrinos within SUSY Grand Unification. }
   
\normalsize\baselineskip=15pt

\section{Introduction}

At present there are many alternative models of neutrino masses. This variety is in part due to the
considerable existing experimental ambiguities. The most crucial questions to be clarified by
experiment are whether the LSND signal will be confirmed or will be excluded and which solar
neutrino solution will eventually be established. If LSND is right we need four light neutrinos, if
not we can do with only the three known ones. Other differences are due to less direct physical
questions like the possible cosmological relevance of neutrinos as hot dark matter. If neutrinos
are an important fraction of the cosmological density, say $\Omega_{\nu}\sim 0.1$, then the average
neutrino mass must be considerably heavier than the splittings that are indicated by the observed
atmospheric and solar oscillation frequencies. For example, for three light neutrinos, only models
with almost degenerate neutrinos, with common mass
$|m_{\nu}|\approx 1~eV$, are compatible with a large hot dark matter component. On the contrary
hierarchical three neutrino models have the largest neutrino mass fixed by $m\approx \sqrt{\Delta
m^2_{atm}}\approx 0.05~ eV$. In most models the smallness of neutrino masses is related to the fact
that $\nu's$ are completely neutral (i.e. they carry no charge which is exactly conserved), they
are Majorana particles and their masses are inversely proportional to the large scale where the
lepton number L conservation is violated. Majorana masses can arise from the see-saw mechanism,
in which case there is some relation with the Dirac masses, or from higher dimension non
renormalisable operators which come from a different sector of the lagrangian density than other
fermion mass terms. 

In my lecture first I will briefly summarise the main categories of
neutrino mass models and give my personal views on them. Then, I will argue in favour of the most
constrained set of models, where there are only three widely split neutrinos, with masses dominated
by the see-saw mechanism and inversely proportional to a large mass close to the Grand Unification
scale
$M_{GUT}$. In this framework neutrino masses are a probe into the physics of GUT's and one can aim
at a comprehensive discussion of all fermion masses. This is for example possible in models based
on $SU(5)\bigotimes U(1)_{flavour}$ or on $SO(10)$ (we always consider SUSY GUT's). This will also
lead us to consider the status of GUT models in view of the experimental bounds on p decay, which
are now very severe also for SUSY models, and of well known naturality problems, like the
doublet-triplet splitting problem. So we will discuss "realistic" as opposed to minimal models,
including a description of the pattern of all fermion masses. We will also mention some recent
ideas on a radically different concept of SUSY
$SU(5)$ where the symmetry is valid in 5 dimensions but is broken by compactification and not by
some Higgs system in the 24 or larger representation. In this version of $SU(5)$ the
doublet-triplet splitting problem is solved elegantly and p decay can naturally be suppressed or
even forbidden by the compactification mechanism. 

This review is in part based on work that I have done
over the recent months with Ferruccio Feruglio and Isabella Masina
\cite{us1}, \cite{us2}, \cite{us3}, \cite{us4}, \cite{us5}, \cite{us6}, \cite{us7}.

\section{Neutrino Masses and Lepton Number Violation}

Neutrino oscillations imply neutrino masses which in turn demand either the existence of
right-handed neutrinos (Dirac masses) or lepton number L violation (Majorana masses) or both.
Given that neutrino masses are certainly extremely small, it is really difficult from the theory
point of view to avoid the conclusion that L must be violated. In fact, it is only in terms of
lepton number violation that the smallness of neutrino masses can be explained as inversely
proportional to the very large scale where L is violated, of order $M_{GUT}$ or even $M_{Planck}$.

Once we accept L violation we gain an elegant explanation for the smallness of neutrino masses
which turn out to be inversely proportional to the large scale where lepton number is violated. If
L is not conserved, even in the absence of
$\nu_R$, Majorana masses can be generated for neutrinos by dimension five operators of the form 
\beq 
O_5=\frac{L^T_i\lambda_{ij}L_jHH}{M}
\label{O5}
\eeq 
with $H$ being the ordinary Higgs doublet, $\lambda$ a matrix in flavour space and M a
large scale of mass, of order $M_{GUT}$ or $M_{Planck}$. Neutrino masses generated by $O_5$ are of
the order
$m_{\nu}\approx v^2/M$ for $\lambda_{ij}\approx {\rm O}(1)$, where $v\sim {\rm O}(100~\GeV)$ is the
vacuum expectation value of the ordinary Higgs. 

We consider that the existence of $\nu_R$ is quite plausible because all GUT groups larger than
SU(5) require them. In particular the fact that $\nu_R$ completes the representation 16 of SO(10):
16=$\bar 5$+10+1, so that all fermions of each family are contained in a single representation of
the unifying group, is too impressive not to be significant. At least as a classification group
SO(10) must be of some relevance. Thus in the following we assume that there are both
$\nu_R$ and lepton number violation. With these assumptions the see-saw mechanism \cite{ss} is
possible which leads to:
\beq m_{\nu}=m_D^T M^{-1}m_D
\eeq That is, the light neutrino masses are
quadratic in the Dirac masses and inversely proportional to the large Majorana mass. Note that for
$m_{\nu}\approx \sqrt{\Delta m^2_{atm}}\approx 0.05~ eV$ and 
$m_{\nu}\approx m_D^2/M$ with $m_D\approx v
\approx 200~GeV$ we find $M\approx 10^{15}~GeV$ which indeed is an impressive indication for
$M_{GUT}$.

If additional non renormalisable terms from $O_5$ are comparatively
non negligible, they should simply be added. After elimination of the heavy right-handed
fields, at the level of the effective low energy theory, the two types of terms are equivalent. In
particular they have identical transformation properties under a chiral change of basis in flavour
space. The difference is, however, that in the see-saw mechanism, the Dirac matrix
$m_D$ is presumably related to ordinary fermion masses because they are both generated by the Higgs
mechanism and both must obey GUT-induced constraints. Thus if we assume the see-saw mechanism more
constraints are implied. In particular we are led to the natural hypothesis that
$m_D$ has a largely dominant third family eigenvalue in analogy to $m_t$, $m_b$ and $m_{\tau}$
which are by far the largest masses among $u$ quarks, $d$ quarks and charged leptons. Once we
accept that $m_D$ is hierarchical it is very difficult to imagine that the effective light neutrino
matrix, generated by the see-saw mechanism, could have eigenvalues very close in absolute value.

\section{Four Neutrino Models}

The LSND signal \cite{sta} has not been confirmed by KARMEN \cite{eit}. It will be soon
double-checked by MiniBoone \cite{sta}. Perhaps it will fade away. But if an oscillation with
$\Delta m^2
\approx 1~ eV^2$ is confirmed then, in presence of three distinct frequencies for LSND, atmospheric
\cite{mcg}, \cite{pul} and solar \cite{tak}, \cite{bel} neutrino oscillations, at least four light
neutrinos are needed. Since LEP has limited to three the number of "active" neutrinos (that is with
weak interactions, or equivalently with non vanishing weak isospin, the only possible gauge charge
of neutrinos) the additional light neutrino(s) must be "sterile", i.e. with vanishing weak isospin.
Note that $\nu_R$ that appears in the see-saw mechanism, if it exists, is a sterile neutrino, but a
heavy one. 

A typical pattern of masses that works for 4-$\nu$ models consists of two pairs of
neutrinos \cite{4nu},\cite{fog} the separation between the two pairs, of order $1~eV$,
corresponding to the LSND frequency. The upper doublet would be almost degerate at $|m|$ of order
$1~eV$ being only split by (the mass difference corresponding to) the atmospheric
$\nu$ frequency, while the lower doublet is split by the solar $\nu$ frequency. This mass
configuration can be compatible with an important fraction of hot dark matter in the universe. A
complication is that the data appear to be incompatible with pure 2-$\nu$ oscillations for
$\nu_e-\nu_s$ oscillations for solar neutrinos and for $\nu_{\mu}-\nu_s$ oscillations for
atmospheric neutrinos (with
$\nu_s$ being a sterile neutrino). There are however viable alternatives. One possibility is
obtained by using the large freedom allowed by the presence of 6 mixing angles in the most general
4-$\nu$ mixing matrix. If 4 angles are significantly different from zero, one can go beyond pure
2-$\nu$ oscillations and, for example, for solar neutrino oscillations
$\nu_e$ can transform into a mixture of $\nu_a~+~\nu_s$, where $\nu_a$ is an active neutrino,
itself a superposition of
$\nu_{\mu}$ and $\nu_{\tau}$ \cite{4nu}. A different alternative is to have many interfering sterile
neutrinos: this is the case in the interesting class of models with extra dimensions, where a whole
tower of Kaluza-Klein neutrinos is introduced. This picture of sterile neutrinos from extra
dimensions is exciting and we now discuss it in some detail \cite{cal}, \cite{moh}.

The context is theories with large extra dimensions. Gravity propagates in all dimensions (bulk),
while SM particles live on a 4-dim brane. As well known \cite{how}, this can make the fundamental
scale of gravity $m_s$ much smaller than the Planck mass $M_P$. In fact, for $d~=~n~+~4$, if $R$ is
the compactification radius we have a geometrical volume factor that suppresses gravity so that:
$(m_sR)^n~=~(M_P/m_s)^2$ and, as a result, $m_s$ can be as small as $\sim 1~TeV$. For neutrino
phenomenology we need a really large extra dimension with $1/R\lappeq0.01~eV$ plus $n-1$ smaller
ones with
$1/\rho\gappeq 1~TeV$. Then we define $m_5$ by $m_5R~=~(M_P/m_s)^2$, or $m_5~=~m_s(m_s\rho)^{n-1}$.
In string theories of gravity there are always scalar fields associated with gravity and their SUSY
fermionic partners (dilatini, modulini). These are particles that propagate in the bulk, have no
gauge interactions and can well play the role of sterile neutrinos. The models based on this
framework \cite{luk} have some good features that make them very appealing at first sight. They
provide a "physical" picture for $\nu_s$. There is a KK tower of recurrences of
$\nu_s$:
\beq
\nu_s(x,y)~=~\frac{1}{\sqrt{R}}\sum_n~\nu_s^{(n)}(x)\cos{\frac{ny}{R}}\label{4nu1}
\eeq with $m_{\nu_s}=n/R$. The tower mixes with the ordinary light active neutrinos in the lepton
doublet L:
\beq L_{mix}~=~h\frac{m_s}{M_P}L\nu_s^{(n)}H\label{4nu2}
\eeq     where H is the Higgs doublet field. Note that the geometrical factor $m_s/M_P$, which
automatically suppresses the Yukawa coupling h, arises naturally from the fact that the sterile
neutrino tower lives in the bulk. Note in passing that $\nu_s$ mixings must be small due to
existing limits from weak processes, supernovae and nucleosynthesis, so that the preferred solution
for 4-$\nu$ models is MSW-(small angle). The interference among a few KK states makes the spectrum
compatible with solar data \cite{cal}, \cite{moh}:
\beq P(\nu_e \rightarrow X)~=~\sum_n\frac{m_e^2}{M_e^2+\frac{n^2}{R^2}}\label{4nu3}
\eeq provided that $1/R\sim 10^{-2}-10^{-3}~eV$ or $R\sim 10^{-3}-10^{-2} ~cm$, that is a really
large extra dimension barely compatible with existing limits \cite{lim}. 

In spite of its good properties there
are problems with this picture, in my opinion. The first property that I do not like of models with
large extra dimensions is that the connection with GUT's is lost. In particular the elegant
explanation of the smallness of neutrino masses in terms of the large scale where the L
conservation is violated in general evaporates. Since $m_s\sim 1~TeV$ is relatively small, what forbids on the
brane an operator of the form
$\frac{1}{m_s} L^T_i\lambda_{ij}L_jHH$ which would lead to by far too large $\nu$ masses? One
must assume L conservation on the brane and that it is only broken by some Majorana masses
of sterile $\nu$'s in the bulk, which I find somewhat ad hoc. Another problem is that we would
expect gravity to know nothing about flavour, but here we would need right-handed partners for
$\nu_e$,
$\nu_{\mu}$ and
$\nu_{\tau}$. Also a single large extra dimension has problems, because it implies \cite{an} a
linear evolution of the gauge couplings with energy from $0.01~eV$ to $m_s\sim 1~TeV$. But more
large extra dimensions lead to
\beq P(\nu_e \rightarrow X)~=~\sum_n\frac{m_e^2}{M_e^2+\frac{n^2}{R^2}}~=~\int dn
n^{d-1}\frac{m_e^2}{M_e^2+\frac{n^2}{R^2}}\label{4nu4}
\eeq For $d>2$ the KK recurrences do not decouple fast enough (the divergence of the integral is
only cut off at
$m_s$) and the mixing becomes very large. Perhaps a compromise at $d=2$ is possible.

In conclusion the models with large extra dimension are interesting because they are speculative
and fascinating but the more conventional framework still appears more plausible at closer
inspection. 

\section{Three Neutrino Models}

We now assume that the LSND signal will not be confirmed, that there are only two distinct neutrino
oscillation frequencies, the atmospheric and the solar frequencies, which can be reproduced with the
known three light neutrino species (for reviews of three neutrino models see \cite{us4},
\cite{barr} where a rather complete set of references can be found).  The two frequencies, are
parametrised in terms of the
$\nu$ mass eigenvalues by 
\beq
\Delta_{sun}\propto m^2_2-m^2_1,~~~~~~~
\Delta_{atm}\propto m^2_3-m^2_{1,2}\label{fre}
\eeq  The numbering 1,2,3 corresponds to our definition of the frequencies and in principle may not
coincide with the family index although this will be the case in the models that we favour. Given
the observed frequencies and  our notation in eq. (\ref{fre}), there are three possible patterns of
mass eigenvalues:
\bea
	 {\rm{Degenerate}}& : & |m_1|\sim |m_2| \sim |m_3|\nonumber\\
	{\rm{Inverted~hierarchy}}& : & |m_1|\sim |m_2| >> |m_3| \nonumber\\
	{\rm{Hierarchical} }& : & |m_3| >> |m_{2,1}|
\label{abc}
\eea We now discuss pro's and con's of the different cases and argue in favour of the hierarchical
option.

\subsection{Degenerate Neutrinos}

At first sight the degenerate case is the most appealing: the observation of nearly maximal
atmospheric neutrino mixing and the possibility that also the solar mixing is large (at present
the MSW-(large angle) solution of the solar neutrino oscillations appears favoured by the data)
suggests that all $\nu$ masses are nearly degenerate. Moreover, the common value of
$|m_{\nu}|$ could be compatible with a large fraction of hot dark matter in the universe for
$|m_{\nu}|\sim 1-2~eV$. In this case, however, the existing limits on the absence of neutrino-less
double beta decay ($0\nu\beta\beta$) imply \cite{GG} double maximal mixing (bimixing) for solar and
atmospheric neutrinos. In fact the quantity which is bound by experiments is the 11 entry of the
$\nu$ mass matrix, which is given by \cite{us4}:
\beq m_{ee}~=~m_1 cos^2\theta_{12}~+~m_2 sin^2\theta_{12}~\lappeq~0.3-0.5~eV\label{3nu1}
\eeq To satisfy this constraint one needs $m_1=-m_2$ (recall that the sign of fermion masses can be
changed by a phase redefinition) and $cos^2\theta_{12}\sim sin^2\theta_{12}$ to a good accuracy (in
fact we need $sin^22\theta_{12} > 0.96$ in order that
$|cos2\theta_{12}|~=~|cos^2\theta_{12}-sin^2\theta_{12}| < 0.2$). Of course this strong constraint
can be relaxed if the common mass is below the hot dark matter maximum. It is true in any case that
a signal of
$0\nu\beta\beta$ near the present limit (like a large relic density of hot dark matter) would be an
indication for nearly degenerate
$\nu$'s.

In general, for naturalness reasons, the splittings cannot be too small with respect to
the common mass, unless there is a protective symmetry \cite{ello}. This is because the wide mass
differences of fermion masses, in particular charged lepton masses, would tend to create neutrino
mass splittings via renormalization group running effects even starting from degenerate masses at a
large scale. For example, the vacuum oscillation solution for solar neutrino oscillations would
imply $\Delta m/m\sim10^{-9}-10^{-11}$ which is difficult to obtain. In this respect the MSW-(large
angle) solution would be favoured, but,  if we insist that $|m_{\nu}|\sim
1-2~eV$, it is not clear that the mixing angle is sufficiently maximal. 

It is clear that in the degenerate case the most likely origin of $\nu$ masses is from dim-5
operators
$O_5=L^T_i\lambda_{ij}L_jHH/M$ and not from the see-saw mechanism $m_{\nu}=m^T_DM^{-1}m_D$.
In fact we expect the $\nu$-Dirac mass $m_D$ to be hierarchical like for all other fermions and a
conspiracy  to reinstaure a
nearly perfect degeneracy between $m_D$ and $M$, which arise from completely different physics,
looks very unplausible. Thus in degenerate models, in general, there is no direct relation with
Dirac masses of quarks and leptons and the possibility of a simultaneous description of all fermion
masses within a grand unified theory is more remote \cite{fri}. 

\subsection{Inverted Hierarchy}

The inverted hierarchy configuration $|m_1|\sim |m_2| >> |m_3|$ consists of two levels $m_1$ and
$m_2$ with small splitting $\Delta m_{12}^2~=~\Delta m_{sun}^2$ and a common mass given by
$m_{1,2}^2\sim \Delta m_{atm}^2\sim 2.5\cdot 10^{-3}~eV^2$ (no large hot dark matter component in this
case). One particularly interesting example of this sort \cite{hall}, which leads to double maximal
mixing, is obtained with the phase choice
$m_1=-m_2$ so that, approximately:
\beq  m_{diag}~=~M [1,-1,0] 
\label{ih1}
\eeq
The effective light neutrino mass matrix
\beq
m_{\nu}~=~Um_{diag}U^T\label{ih2}
\eeq
which corresponds to the mixing matrix of double maximal mixing $c=s=1/\sqrt{2}$:   
\beq  U_{fi}= 
\left[\matrix{ c&-s&0 \cr s/\sqrt{2}&c/\sqrt{2}&-1/\sqrt{2}\cr s/\sqrt{2}&c/\sqrt{2}&+1/\sqrt{2}     } 
\right] ~~~~~.
\label{ufi}
\eeq
is given by:
\beq  m_{\nu}~=~\frac{M}{\sqrt{2}} 
\left[\matrix{ 0&1&1 \cr 1&0&0\cr
1&0&0     } 
\right] ~~~~~.
\label{ih3}
\eeq
The structure of $m_{\nu}$ can be reproduced by imposing a flavour symmetry $L_e-L_{\mu}-L_{\tau}$
starting from
$O_5=L^T_i\lambda_{ij}L_jHH/M$. The $1-2$ degeneracy remains stable under radiative
corrections. The preferred solar solutions are vacuum oscillations or the LOW solution. The
MSW-(large angle) could be also compatible if the mixing angle is large enough. The required
dominance of $O_5$ leads to the same comments as the degenerate models of the previous section.

\subsection{Hierarchical}

We now discuss the class of models which we consider of particular interest because this is the
most constrained framework which allows a comprehensive combined study of all fermion
masses in GUT's. We assume three widely split $\nu$'s and the existence of a right-handed neutrino
for each generation, as required to complete a 16-dim representation of $SO(10)$ for each
generation. We then assume dominance of the see-saw
mechanism
$m_{\nu}=m_D^TM^{-1}m_D$. We know that the third-generation eigenvalue of the Dirac mass matrices
of up and down quarks and of charged leptons is systematically the largest one. It is natural
to imagine that this property will also be true for the Dirac mass of $\nu$'s: $diag[m_D]\sim
[0,0,m_{D3}]$. After see-saw we expect $m_{\nu}$ to be
even more hierarchical being quadratic in
$m_D$ (barring fine-tuned compensations between $m_D$ and $M$). The amount of hierarchy,
$m^2_3/m^2_2=\Delta m^2_{atm} /\Delta m^2_{sun}$, depends on which solar neutrino solution is
adopted: the hierarchy is maximal for vacuum oscillations and LOW solutions, is moderate for MSW in
general and could become quite mild for the upper $\Delta m^2$ domain of the MSW-(large angle)
solution. A possible difficulty is that one is used to expect that large splittings correspond
to small mixings because normally only close-by states are strongly mixed. In a 2 by 2 matrix
context the requirement of large splitting and large mixings leads to a condition of vanishing
determinant. For example the matrix
\beq m\propto 
\left[\matrix{ x^2&x\cr x&1    } 
\right]~~~~~. 
\label{fourteen}
\eeq 
has eigenvalues 0 and $1+x^2$ and for $x$ of 0(1) the mixing is large. Thus in the limit of
neglecting small mass terms of order $m_{1,2}$ the demands of large atmospheric neutrino mixing and
dominance of $m_3$ translate into the condition that the 2 by 2 subdeterminant 23 of the 3 by 3
mixing matrix approximately vanishes. The problem is to show that this vanishing can be arranged in
a natural way without fine tuning. Once near maximal atmospheric neutrino mixing is reproduced
the solar neutrino mixing can be arranged to be either small of large without difficulty by
implementing suitable relations among the small mass terms. 

It is not difficult to imagine mechanisms that naturally lead to the approximate vanishing of the
23 sub-determinant. For example \cite{hall}, \cite{king}, assume that one $\nu_R$ is particularly
light and coupled to
$\mu$ and $\tau$. In a 2 by 2 simplified context if we have
\beq M\propto 
\left[\matrix{ \epsilon&0\cr 0&1    } 
\right];~~~~~ M^{-1}\approx\left[\matrix{ 1/\epsilon&0\cr 0&0    } 
\right]
\label{md0}
\eeq then for a generic $m_D$ we find
\beq m_{\nu}~=~m_D^TM^{-1}M_D\sim 
\left[\matrix{ a&c\cr b&d   } 
\right] \left[\matrix{ 1/\epsilon&0\cr 0&0    } 
\right]\left[\matrix{ a&b\cr c&d   }\right]~=~\frac{1}{\epsilon}\left[\matrix{ a^2&ac\cr ac&c^2   } 
\right] 
\label{md1}
\eeq
A different possibility that we find attractive is that, in the limit of neglecting terms of order
$m_{1,2}$ and, in the basis where charged leptons are diagonal, the Dirac matrix $m_D$,
defined by $\bar R m_D L$, takes the approximate form:
\beq m_D\propto 
\left[\matrix{ 0&0&0\cr 0&0&0\cr 0&x&1    } 
\right]~~~~~. 
\label{md00}
\eeq This matrix has the property that for a generic Majorana matrix $M$ one finds:
\beq m_{\nu}=m^T_D M^{-1}m_D\propto 
\left[\matrix{ 0&0&0\cr 0&x^2&x\cr 0&x&1    } 
\right]~~~~~. 
\label{mn0}
\eeq The only condition on $M^{-1}$ is that the 33 entry is non zero. But when the approximately
vanishing matrix elements are replaced by small terms, one must also assume that no new $o(1)$ terms
are generated in $m_{\nu}$ by a compensation between small terms in $m_D$ and large terms in
$M$. It is important for the following discussion to observe that
$m_D$ given by eq. (\ref{md00}) under a change of basis transforms as $m'_D->V^{\dagger} m_D U$
where V and U rotate the right and left fields respectively. It is easy to check that in order to
make $m_D$ diagonal we need large left mixings (i.e. large off diagonal terms in the matrix that
rotates left-handed fields).
Thus the question is how to reconcile large left-handed mixings
in the leptonic sector with the observed near diagonal form of $V_{CKM}$, the quark mixing matrix.
Strictly speaking, since $V_{CKM}=U^{\dagger}_u U_d$, the individual matrices $U_u$ and $U_d$ need
not be near diagonal, but
$V_{CKM}$ does, while the analogue for leptons apparently cannot be near diagonal. However nothing
forbids for quarks that, in the basis where $m_u$ is diagonal, the $d$ quark matrix has large non
diagonal terms that can be rotated away by a pure right-handed rotation. We suggest that this is so
and that in some way right-handed mixings for quarks correspond to left-handed mixings for leptons.

In the context of (Susy) SU(5) there is a very attractive hint of how the present
mechanism can be realized. In the
$\bar 5$ of SU(5) the $d^c$ singlet appears together with the lepton doublet $(\nu,e)$. The $(u,d)$
doublet and $e^c$ belong to the 10 and $\nu^c$ to the 1 and similarly for the other families. As a
consequence, in the simplest model with mass terms arising from only Higgs pentaplets, the Dirac
matrix of down quarks is the transpose of the charged lepton matrix:
$m^d_D=(m^l_D)^T$. Thus, indeed, a large mixing for right-handed down quarks corresponds to a large
left-handed mixing for charged leptons.  At leading order we may have:
\beq m_d=(m_l)^T=
\left[
\matrix{ 0&0&0\cr 0&0&x\cr 0&0&1}
\right]v_d
\eeq In the same simplest approximation with  5 or $\bar 5$ Higgs, the up quark mass matrix is
symmetric, so that left and right mixing matrices are equal in this case. Then small mixings for up
quarks and small left-handed mixings for down quarks are sufficient to guarantee small $V_{CKM}$
mixing angles even for large $d$ quark right-handed mixings.  If these small mixings are neglected,
we expect:
\beq m_u=
\left[
\matrix{ 0&0&0\cr 0&0&0\cr 0&0&1}
\right]v_u
\eeq When the charged lepton matrix is diagonalized the large left-handed mixing of the charged
leptons is transferred to the neutrinos. Note that in SU(5) we can diagonalize the $u$ mass matrix
by a rotation of the fields in the 10, the Majorana matrix $M$ by a rotation of the 1 and the
effective light neutrino matrix
$m_\nu$ by a rotation of the $\bar 5$. In this basis the $d$ quark mass matrix fixes $V_{CKM}$ and
the charged lepton mass matrix fixes neutrino mixings. It is well known that a model where the down
and the charged lepton matrices are exactly the transpose of one another cannot be exactly true
because of the $e/d$ and
$\mu/s$ mass ratios. It is also known that one remedy to this problem is to add some Higgs
component in the 45 representation of SU(5) \cite{jg}. A different kind of solution \cite{eg} will
be described later. But the symmetry under transposition can still be a good guideline if we are
only interested in the order of magnitude of the matrix entries and not in their exact values.
Similarly, the Dirac neutrino mass matrix
$m_D$ is the same as the up quark mass matrix in the very crude model where the Higgs pentaplets
come from a pure 10 representation of SO(10):
$m_D=m_u$. For $m_D$ the dominance of the third family eigenvalue  as well as a near diagonal
form could be an order of magnitude remnant of this broken symmetry. Thus, neglecting small terms,
the neutrino Dirac matrix in the basis where charged leptons are diagonal could be directly
obtained in the form of eq. (\ref{md00}).

\section{ Simple Examples with Horizontal Abelian Charges}
 
We discuss here some explicit examples of the mechanism under discussion in the framework
of a unified Susy $SU(5)$ theory with an additional 
$U(1)_F$ flavour symmetry \cite{fro}. If, for a given interaction vertex, the $U(1)_F$ charges do
not add to zero, the vertex is forbidden in the symmetry limit. But the symmetry is spontaneously
broken by the vev $v_f$ of a number of "flavon" fields with non vanishing charge. Then a forbidden
coupling is rescued but is suppressed by powers of the small parameters $v_f/M$ with the exponent
larger for larger charge mismatch. We expect $v_f\gappeq M_{GUT}$ and $M \lappeq M_P$. Here we
discuss some aspects of the description of fermion masses in these models. In the following
sections we will consider how to imbed these concepts within more complete and realistic $SU(5)$
models. We will also discuss the need and the options to go beyond minimal models.

In these models the known generations of quarks and
leptons are contained in triplets
$\Psi_{10}^a$ and
$\Psi_{\bar 5}^a$, $(a=1,2,3)$ transforming as $10$ and ${\bar 5}$ of $SU(5)$, respectively. Three
more $SU(5)$ singlets
$\Psi_1^a$ describe the right-handed neutrinos. In SUSY models we have two Higgs multiplets, which
transform as 5 and $\bar 5$ in the minimal model. We first assume that they have the same charge.
The simplest models are obtained by allowing all the third generation masses already in the
symmetric limit. This is realised by taking vanishing charges for the Higgses and for the third
generation components $\Psi_{10}^3$, $\Psi_{\bar 5}^3$ and $\Psi_1^3$. We can arrange the unit of
charge in such a way that the Cabibbo angle, which we consider as the typical hierachy parameter of
fermion masses and mixings, is obtained when the suppression exponent is unity. Remember that the
Cabibbo angle is not too small, $\lambda \sim 0.22$ and that in $U(1)_F$ models all mass matrix
elements are of the form of a power of a suppression factor times a number of order unity, so that
only their order of suppression is defined. As a consequence, in practice,  we can limit ourselves
to integral charges in our units, for simplicity (for example, $\sqrt{\lambda} \sim 1/2$ is already
almost unsuppressed). 

After these preliminaries let's first try a simplest model with all charges being non negative and
containing one single flavon of negative charge. For example, we could take \cite{buch} (see also
\cite{abelian})
\bea
\Psi_{10}     & \sim & (4,2,0) \label{twentyone}\\
\Psi_{\bar 5} & \sim & (2,0,0) \label{twentytwo}\\
\Psi_1        & \sim & (4,2,0) \label{twentythree}
\eea       
In this case a typical mass matrix has the form
\beq m~=~
\left[
\matrix{
y_{11}\lambda^{q_1+q'_1}&y_{12}\lambda^{q_1+q'_2}&y_{13}\lambda^{q_1+q'_3}\cr
y_{21}\lambda^{q_2+q'_1}&y_{22}\lambda^{q_2+q'_2}&y_{23}\lambda^{q_2+q'_3}\cr
y_{31}\lambda^{q_3+q'_1}&y_{32}\lambda^{q_3+q'_2}&y_{33}\lambda^{q_3+q'_3}}
\right]v\label{twentyfour}
\eeq 
where all the $y{ij}$ are of order 1 and $q_i$ and $q'_i$ are the charges of 10,10 for $m_u$, of
$\bar 5$,10 for $m_d$ or $m^T_l$, of 1,$\bar 5$ for $m_D$ (the Dirac $\nu$ mass), and of 1,1
for M, the RR Majorana $\nu$ mass. Note the two vanishing charges in $\Psi_{\bar 5}$. They are
essential for this mechanism: for example they imply that the 32, 33 matrix elements of
$m_D$ are of order 1. It is important to observe that
$m$ can be written as:
\beq
m~=~\lambda^q y \lambda^{q'}\label{m1}
\eeq   
where $\lambda_q=diag[\lambda_{q_1},\lambda_{q_2},\lambda_{q_3}]$ and y is the $y_{ij}$ matrix. As a
consequence when we start from the Dirac $\nu$ matrix: $m_D~=~\lambda^{q_{1}} y_D \lambda^{q_{\bar
5}}$ and the RR Majorana matrix $M~=~\lambda^{q_{1}} y_M \lambda^{q_{1}}$ and write down the see-saw
expression for $m_{\nu}=m_D^T M^{-1} m_D$, we find that the dependence on the $q_{1}$ charges drops
out and only that from $q_{\bar5}$ remains. On the one hand this is good because it corresponds to
the fact that the effective light neutrino Majorana mass matrix $m_{\nu}\sim L^TL$ can be
written  in terms of
$q_{\bar5}$ only. In particular the 22,23,32,33 matrix elements of $m_{\nu}$ are of order 1, which
implies large mixings in the 23 sector. On the other hand the sub determinant 23 is not suppressed
in this case, so that the splitting between the 2 and 3 light neutrino masses is in general small.
In spite of the fact that $m_D$ is, in first approximation, of the form in eq. (\ref{md00}) the
strong correlations between
$m_D$ and $M$ implied by the simple charge structure of the model destroy the vanishing of the 23
sub determinant that would be guaranteed for generic $M$. Models of this sort have been proposed in
the literature \cite{buch}, \cite{abelian}. The hierarchy between
$m_2$ and
$m_3$ is considered accidental and better be moderate. The preferred solar solution in this case is
MSW-(small angle) because if $m_1$ is suppressed the solar mixing angle is typically small.

Models with natural large 23 splittings are obtained if we allow negative charges and, at the same
time, either introduce flavons of opposite charges or stipulate that matrix elements with overall
negative charge are put to zero. We now discuss a model of this sort \cite{us3}. We assign to the
fermion fields the set of
$F$-charges given by:
\bea
\Psi_{10}     & \sim & (3,2,0) \label{c10}\\
\Psi_{\bar 5} & \sim & (3,0,0) \label{c5b}\\
\Psi_1        & \sim & (1,-1,0) \label{c1}
\eea   We consider the Yukawa coupling allowed by $U(1)_F$-neutral  Higgs multiplets
$\varphi_5$ and
$\varphi_{\bar5}$ in the $5$ and ${\bar 5}$ $SU(5)$ representations and by a pair $\theta$ and
${\bar\theta}$ of $SU(5)$ singlets with $F=1$ and $F=-1$, respectively. 

In the quark sector we obtain :
\beq m_u=(m_u)^T=
\left[
\matrix{
\lambda^6&\lambda^5&\lambda^3\cr
\lambda^5&\lambda^4&\lambda^2\cr
\lambda^3&\lambda^2&1}
\right]v_u~~,~~~~~~~ m_d=
\left[
\matrix{
\lambda^6&\lambda^5&\lambda^3\cr
\lambda^3&\lambda^2&1\cr
\lambda^3&\lambda^2&1}
\right]v_d~~,
\label{mquark}
\eeq  from which we get for the eigenvalues the order-of-magnitude relations:
\bea m_u:m_c:m_t & = &\lambda^6:\lambda^4:1 \nonumber\\ m_d:m_s:m_b & = &\lambda^6:\lambda^2:1
\eea and 
\beq V_{us}\sim \lambda~,~~~~~ V_{ub}\sim \lambda^3~,~~~~~ V_{cb}\sim \lambda^2~.
\eeq 

Here $v_u\equiv\langle \varphi_5 \rangle$, 
$v_d\equiv\langle \varphi_{\bar 5} \rangle$  and $\lambda$, arising from the $\bar{\theta}$ vev,
is, as above, of the order of the Cabibbo angle. For non-negative
$F$-charges, the elements of  the quark mixing matrix $V_{CKM}$ depend only on the charge
differences  of the left-handed quark doublet \cite{fro}. Up to a constant shift, this defines the
choice in eq. (\ref{c10}). Equal $F$-charges for $\Psi_{\bar 5}^{2,3}$ (see eq. (\ref{c5b})) are
then required to fit
$m_b$ and $m_s$. We will comment on the lightest quark masses later on.

At this level, the mass matrix for the charged leptons  is the transpose of $m_d$:
\beq m_l=(m_d)^T
\eeq and we find:
\beq m_e:m_\mu:m_\tau  = \lambda^6:\lambda^2:1 
\eeq The O(1) off-diagonal entry of $m_l$ gives rise to a large left-handed  mixing in the 23
block which corresponds to a large right-handed mixing in the $d$ mass matrix. In the neutrino
sector, the Dirac and Majorana mass matrices are given by:
\beq m_D=
\left[
\matrix{
\lambda^4&\lambda&\lambda\cr
\lambda^2&\lambda'&\lambda'\cr
\lambda^3&1&1}
\right]v_u~~,~~~~~~~~ M=
\left[
\matrix{
\lambda^2&1&\lambda\cr 1&\lambda'^2&\lambda'\cr
\lambda&\lambda'&1}
\right]{\bar M}~~,
\eeq where $\lambda'$ is related to $\theta$ and ${\bar M}$  denotes the large mass scale
associated to the right-handed neutrinos: ${\bar M}\gg v_{u,d}$.

After diagonalization of the charged lepton sector and after integrating out the heavy right-handed
neutrinos we obtain the following neutrino mass matrix in the low-energy effective theory:
\beq m_\nu=
\left[
\matrix{
\lambda^6&\lambda^3&\lambda^3\cr
\lambda^3&1&1\cr
\lambda^3&1&1}\right]{v_u^2\over {\bar M}}
\label{mnu}
\eeq where we have taken $\lambda\sim\lambda'$. The O(1) elements in the 23 block are produced by
combining the large  left-handed mixing induced by the charged lepton sector and the large
left-handed mixing in $m_D$. A crucial property of
$m_\nu$ is that, as a result of the sea-saw mechanism and of the specific $U(1)_F$ charge
assignment, the determinant of the 23 block is $\underline{automatically}$ of $O(\lambda^2)$ (for
this the presence of negative charge values, leading to the presence of both $\lambda$ and
$\lambda'$ is essential \cite{us2}).

It is easy to verify that the eigenvalues of $m_\nu$ satisfy  the relations:
\beq m_1:m_2:m_3  = \lambda^4:\lambda^2:1~~.
\eeq The atmospheric neutrino oscillations require 
$m_3^2\sim 10^{-3}~{\rm eV}^2$. From eq. (\ref{mnu}), taking $v_u\sim 250~{\rm GeV}$, the mass
scale ${\bar M}$ of the heavy Majorana neutrinos turns out to be close to the unification scale, 
${\bar M}\sim 10^{15}~{\rm GeV}$. The squared mass difference between the lightest states is  of
$O(\lambda^4)~m_3^2$, appropriate to the MSW solution  to the solar neutrino problem. Finally,
beyond the large mixing in the 23 sector,
$m_\nu$  provides a mixing angle $s \sim (\lambda/2)$ in the 12 sector, close to the range
preferred by the small angle MSW solution. In general $U_{e3}$ is non-vanishing, of $O(\lambda^3)$.

In general, the charge assignment under 
$U(1)_F$ allows for non-canonical kinetic terms that represent an additional source of mixing. Such
terms are allowed by the underlying flavour symmetry and it would be unnatural to tune them to the
canonical form. The results quoted up to now  remain unchanged after including the effects related
to the most general kinetic terms, via appropriate  rotations and rescaling in the flavour space. 

Obviously, the order of magnitude description offered by this model is not intended to account for
all the details of fermion masses. Even neglecting the parameters associated with the $CP$
violating observables, some of the relevant observables are somewhat marginally reproduced.   For
instance we obtain $m_u/m_t\sim \lambda^6$  which is perhaps too large. However we find it
remarkable that in such a simple scheme most of the 12 independent  fermion masses and the 6 mixing
angles turn out to have the correct order of magnitude. Notice also that this model prefers large
values of
$\tan\beta\equiv v_u/v_d$. This is a consequence of the equality $F(\Psi_{10}^3)=F(\Psi_{\bar
5}^3)$ (see eqs. (\ref{c10}) and (\ref{c5b})). In this case the Yukawa couplings of top and bottom
quarks  are expected to be of the same order of magnitude, while the large
$m_t/m_b$ ratio is attributed to $v_u \gg  v_d$ (there may be factors O(1) modifying these
considerations, of course).  Alternatively, to keep
$\tan\beta$ small, one could  suppress $m_b/m_t$ by adopting different $F$-charges for the 
$\Psi_{\bar 5}^3$ and $\Psi_{10}^3$ or for the $5$ and $\bar 5$ Higgs, as we will see in the next
section.

A common problem of all $SU(5)$ unified theories based on a minimal higgs structure is represented
by the relation
$m_l=(m_d)^T$ that, while leading to the successful $m_b=m_\tau$ boundary condition at the GUT
scale, provides the wrong prediction
$m_d/m_s=m_e/m_\mu$ (which, however, is an acceptable order of magnitude equality). We can easily
overcome this problem and improve the picture \cite{eg} by introducing an additional supermultiplet
${\bar\theta}_{24}$ transforming in the adjoint representation of $SU(5)$ and  possessing a
negative $U(1)_F$ charge,
$-n~~(n>0)$.  Under these conditions, a positive
$F$-charge $f$ carried by the matrix elements 
$\Psi_{10}^a \Psi_{\bar 5}^b$ can be compensated  in several different ways by monomials of the kind
$({\bar\theta})^p({\bar\theta}_{24})^q$, with
$p+n q=f$. Each of these possibilities represents an independent contribution to the down quark and
charged lepton mass matrices, occurring with an unknown coefficient of O(1). Moreover the product
$({\bar\theta}_{24})^q \varphi_{\bar 5}$ contains both the ${\bar 5}$ and the $\overline{45}$
$SU(5)$ representations,  allowing for a differentiation between the down quarks and the charged
leptons. The only, welcome, exceptions are  given by the O(1) entries that do not require any
compensation and, at the leading order, remain the same for charged leptons and  down quarks. This
preserves the good
$m_b=m_\tau$ prediction. Since a perturbation of O(1) in the subleading matrix elements  is
sufficient to cure the bad
$m_d/m_s=m_e/m_\mu$ relation, we can safely assume that 
$\langle{\bar\theta}_{24}\rangle/M_P\sim\lambda^n$, to preserve the correct order-of-magnitude
predictions in the remaining sectors. 

A general problem common to all models dealing with flavour  is that of recovering the correct 
vacuum structure by minimizing the effective potential of the theory. It may be noticed that the
presence of two multiplets $\theta$ and
${\bar \theta}$ with opposite $F$ charges could hardly be reconciled, without adding extra
structure to the model, with a large common VEV for these fields, due to possible analytic terms of
the kind $(\theta {\bar \theta})^n$ in the superpotential. We find therefore instructive
to explore the consequences of allowing only the negatively charged ${\bar \theta}$
field in the theory.

It can be immediately recognized that, while the quark mass matrices of eqs. (\ref{mquark}) are
unchanged, in the neutrino sector the Dirac  and Majorana matrices get modified into:
\beq m_D=
\left[
\matrix{
\lambda^4&\lambda&\lambda\cr
\lambda^2&0&0\cr
\lambda^3&1&1}
\right]v_u~~,~~~~~~~~ M=
\left[
\matrix{
\lambda^2&1&\lambda\cr 1&0&0\cr
\lambda&0&1}
\right]{\bar M}~~.
\eeq The zeros are due to the analytic property of the superpotential that makes impossible to form
the corresponding $F$ invariant by using ${\bar \theta}$ alone. These zeros should not be taken
literally, as they will be eventually   filled by small terms coming, for instance, from the
diagonalization of the charged lepton mass matrix and from the transformation that put the kinetic
terms into canonical form. It is however interesting to work out, in first approximation, the case 
of exactly zero entries in $m_D$ and $M$, when forbidden by $F$.

The neutrino mass matrix obtained via see-saw from $m_D$ and $M$ has the same pattern as the one
displayed in eq. (\ref{mnu}). A closer inspection reveals that the determinant of the 23 block is
identically zero, independently from
$\lambda$. This leads to the following pattern of masses:
\beq m_1:m_2:m_3  = \lambda^3:\lambda^3:1~~,~~~~~m_1^2-m_2^2 = {\rm O}(\lambda^9)~~.
\eeq Moreover the mixing in the 12 sector is almost maximal:
\beq {s\over c}={\pi\over 4}+{\rm O}(\lambda^3)~~.
\eeq For $\lambda\sim 0.2$, both the squared mass difference $(m_1^2-m_2^2)/m_3^2$  and $\sin^2
2\theta_{sun}$ are remarkably close to the values  required by the vacuum oscillation solution to
the solar neutrino problem. This property  remains reasonably stable against the perturbations
induced by small terms (of order $\lambda^5$) replacing the zeros, coming from the diagonalization
of the charged lepton sector  and by the transformations that render the kinetic terms canonical.
We find quite interesting that also the just-so solution, requiring  an intriguingly small mass
difference and a bimaximal mixing, can be reproduced, at least at the level of order of magnitudes,
in the context of a "minimal" model of flavour compatible with supersymmetric SU(5). In this case
the role played by supersymmetry  is essential, a non-supersymmetric model with ${\bar \theta}$
alone  not being distinguishable from the version with both
$\theta$ and ${\bar \theta}$, as far as low-energy flavour properties are concerned.

\section{From Minimal to Realistic SUSY SU(5)}

In this section, following the lines of a recent study \cite{us6}, we address the question
whether the smallest SUSY SU(5) symmetry group can still be considered as a basis for a
realistic GUT model. The minimal model has large fine tuning problems (e.g. the doublet-triplet
splitting problem) and phenomenological problems from the new improved limits on proton decay
\cite{skp}. Also, analyses of particular aspects of GUT's often leave aside the problem of embedding
the sector under discussion into a consistent whole. So the problem arises of going beyond minimal
toy models by formulating sufficiently realistic, not unnecessarily complicated, relatively
complete models that can serve as benchmarks to be compared with experiment. More appropriately,
instead of "realistic" we should say "not grossly unrealistic" because it is clear that many
important details cannot be sufficiently controlled and assumptions must be made.   The model we
aim at should not rely on large fine tunings and must lead to an acceptable phenomenology. This
includes coupling unification with an acceptable value of
$\alpha_s(m_Z)$, given $\alpha$ and
$sin^2\theta_W$ at $m_Z$, compatibility with the bounds on proton decay, agreement with the
observed fermion mass spectrum, also considering neutrino masses and mixings and so on. The success
or failure of the programme of constructing realistic models can decide whether or not a stage of
gauge unification is a likely possibility. 

We
indeed have presented in ref.\cite{us6} an explicit example of a "realistic" SU(5) model, which uses
a
$U(1)_F$ symmetry as a crucial ingredient. In this model the doublet-triplet splitting problem is
solved by the missing partner mechanism \cite{mp} stabilised by the flavour symmetry against the
occurrence of doublet mass lifting due to non renormalisable operators. Relatively large
representations (50,
$\bar{50}$, 75) have to be introduced for this purpose. A good effect of this proliferation of
states is that the value of
$\alpha_s(m_Z)$ obtained from coupling unification in the next to the leading order perturbative
approximation receives important negative corrections from threshold effects near the GUT scale
arising from mass splittings inside the 75. As a result, the central value changes from 
$\alpha_s(m_Z)\approx 0.129$ in minimal SUSY SU(5) down to
$\alpha_s(m_Z)\approx 0.116$, in better agreement with observation. At the same time, an increase of
the effective mass that mediates proton decay by a factor of typically 20-30 is obtained to
optimize the value of $\alpha_s(m_Z)$. So finally the value of the strong coupling is in
better agreement with the experimental value and the proton decay rate is smaller by a factor
400-1000 than in the minimal model (in addition the rigid relation of the minimal model between
mass terms and proton decay amplitudes is released, so that the rate can further be reduced) . The
presence of these large representations also has the consequence that the asymptotic freedom of
SU(5) is spoiled and the associated gauge coupling becomes non perturbative below 
$M_P$. We argue that this property far from being unacceptable can actually
be useful to obtain better results for fermion masses and proton decay. The same $U(1)_F$ flavour
symmetry that stabilizes the missing partner mechanism explains the hierarchical structure of
fermion masses. In the neutrino sector, mass matrices similar to those discussed in the previous
section are obtained. In the
present particular version maximal mixing also for solar neutrinos is preferred.

While we refer to the original paper for a complete discussion, here we only summarise the fermion
mass sector of the model, which is of relevance for neutrinos.
At variance with the previous models we adopt in this case different $U(1)_F$ charges
for the Higgs field
$H\sim 5$ and
$\bar H\sim
\bar 5$:
\beq F(H)=-2~~\rm{and}~~F(\bar H)=1,\label{7a}  
\eeq For matter fields
\bea F(\Psi_{10})=(4,3,1)\nonumber \\ F(\Psi_{\bar 5})=(5,2,2)\nonumber \\ 
F(\Psi_{1})=(1,-1,0)\label{8} 
\eea The Yukawa mass matrices are, in first approximation, of the form:
\beq m_u~=~ 
\left[\matrix{
\lambda^6&\lambda^5&\lambda^3\cr
\lambda^5&\lambda^4&\lambda^2\cr
\lambda^3&\lambda^2&1    } 
\right]~v_u/\sqrt{2}~~~~, 
\label{mua}
\eeq
\beq m_d~=~ 
\left[\matrix{
\lambda^6&\lambda^5&\lambda^3\cr
\lambda^3&\lambda^2&1\cr
\lambda^3&\lambda^2&1    } 
\right]~v_d\lambda^4/\sqrt{2}~~~~=m_l^T, 
\label{mda}
\eeq
\beq m_{\nu}~=~ 
\left[\matrix{
\lambda^4&\lambda&\lambda\cr
\lambda^2&0&0\cr
\lambda^3&1&1    } 
\right]~v_u/\sqrt{2}~~~~, 
\label{mnua}
\eeq
\beq m_{maj}~=~ 
\left[\matrix{
\lambda^2&1&\lambda\cr 1&0&0\cr
\lambda&0&1    } 
\right]~M~~~~, 
\label{Ma}
\eeq For a correct first approximation of the observed spectrum we need
$\lambda\approx\lambda_C\approx 0.22$,
$\lambda_C$ being the Cabibbo angle. These mass matrices closely match those of
the previous section, with two important special features. First, we have here that
$tan\beta=v_u/v_d\approx m_t/m_b\lambda^4$, which is small. The factor $\lambda^4$ is obtained
as a consequence of the Higgs and matter fields charges F, while previously the $H$ and
$\bar H$ charges were taken as zero. We recall that a value of $\tan{\beta}$ near 1 is an
advantage for suppressing proton decay. Of course the limits from LEP that indicate that
$tan\beta \gappeq 2-3$ must be and can be easily taken into account. Second, the zero entries in the
mass matrices of the neutrino sector occur because the negatively F-charged flavon fields have no
counterpart with positive F-charge in this model. Neglected small effects could partially fill up
the zeroes. As already explained these zeroes lead to near maximal mixing also for solar
neutrinos. A problematic aspect of this zeroth order approximation to the mass matrices is the
relation
$m_d=m_l^T$. The necessary corrective terms can
arise from the neglected  higher order terms from non renormalisable operators
with the insertion of n factors of the 75, which break the transposition relation between $m_d$
and $m_l$. With reasonable values of the coefficients of order 1 we obtain double nearly maximal
mixing and $\theta_{13}\sim 0.05$. The preferred solar solutions are LOW or vacuum oscillations.

\section{SU(5) Unification in Extra Dimensions}

Recently it has been observed that the GUT
gauge symmetry could be actually realized in 5 (or more) space-time dimensions and broken down to
the the Standard Model (SM) by compactification
\footnote{Grand unified supersymmetric models in six dimensions, with the grand unified scale
related to the compactification scale were also proposed by Fayet \cite{faye}.}. In particular a
model with N=2 Supersymmetry (SUSY) and gauge SU(5) in 5 dimensions has been proposed \cite{kawa} 
where the GUT symmetry is broken by compactification on
$S^1/(Z_2\times Z_2')$ down to a N=1 SUSY-extended version of the SM on a 4-dimensional brane.  In
this model many good properties of GUT's, like coupling unification and charge quantization are
maintained while some unsatisfactory properties of the conventional breaking mechanism, like
doublet-triplet splitting, are avoided. In a recent paper of ours \cite {us7} we have elaborated
further on this class of models. We differ from ref. \cite{kawa} (and also from the later reference
\cite{hn}) in the form of the interactions on the 4-dimensional brane. As a consequence we not only
avoid the problem of the doublet-triplet splitting but also directly suppress or even forbid proton
decay, since the conventional higgsino and gauge boson exchange amplitudes are absent, as a
consequence of
$Z_2\times Z_2'$ parity assignments on matter fields on the brane.   Most good predictions of SUSY
SU(5) are thus maintained without unnatural fine tunings as needed in the minimal model. We find
that the relations among fermion masses implied by the minimal model, for example $m_b=m_{\tau}$ at
$M_{GUT}$ are preserved in our version of the model, although the Yukawa interactions are not
fully SU(5) symmetric. The mechanism that forbids proton decay still allows Majorana mass terms
for neutrinos so that the good potentiality of SU(5) for the description of neutrino masses and
mixing is preserved. This class of models offers a new perspective on how the GUT symmetry and
symmetry-breaking could be realized.

\section{SO(10) Models}

Models based on $SO(10)$ times a flavour symmetry are more difficult to construct because a whole
generation is contained in the 16, so that, for example for $U(1)_F$, one would have the same value
of the charge for all quarks and leptons of each generation, which is too rigid. But the mechanism
discussed sofar, based on asymmetric mass matrices, can be embedded in an
$SO(10)$ grand-unified theory in a rather economic way
\cite{alb}, \cite{barr}. The 33 entries of the fermion mass matrices can be obtained through the
coupling
${\bf 16}_3 {\bf 16}_3 {\bf 10}_H$ among the fermions in the third generation, ${\bf 16}_3$, and a
Higgs tenplet ${\bf 10}_H$. The two independent VEVs of the tenplet $v_u$ and $v_d$ give mass,
respectively, to $t/\nu_\tau$ and $b/\tau$. The keypoint to obtain an asymmetric texture is the
introduction of an operator of the kind ${\bf 16}_2 {\bf 16}_H {\bf 16}_3 {\bf 16}_H'$ . This
operator is thought to arise by integrating out an heavy {\bf 10} that couples both to ${\bf 16}_2
{\bf 16}_H$ and to ${\bf 16}_3 {\bf 16}_H'$. If the ${\bf 16}_H$ develops a VEV breaking $SO(10)$
down to $SU(5)$ at a large scale, then, in terms of
$SU(5)$ representations, we get an effective coupling of the kind ${\bf \bar{5}}_2 {\bf 10}_3
{\bf\bar{5}}_H$, with a coefficient that can be of order one. This coupling contributes to the 23
entry of the down quark mass matrix  and to the 32 entry of the charged lepton mass matrix,
realizing the desired asymmetry.   To distinguish the lepton and quark sectors one can further
introduce  an operator of the form ${\bf 16}_i {\bf 16}_j {\bf 10}_H {\bf 45}_H$, $(i,j=2,3)$, with
the VEV of the 
${\bf 45}_H$ pointing in the $B-L$ direction. Additional operators, still of the type 
${\bf 16}_i {\bf 16}_j {\bf 16}_H {\bf 16}_H'$ can contribute to the matrix elements of the first
generation. The mass matrices look like:
\beq m_u=
\left[
\matrix{ 0& 0& 0\cr 0& 0& \epsilon/3\cr 0&-\epsilon/3&1}
\right]v_u~~,~~~~~~~ m_d=
\left[
\matrix{ 0&\delta&\delta'\cr
\delta&0&\sigma+\epsilon/3\cr
\delta'&-\epsilon/3&1}
\right]v_d~~,
\label{mquark1}
\eeq 
\beq m_D=
\left[
\matrix{ 0& 0& 0\cr 0& 0& -\epsilon\cr 0& \epsilon&1}
\right]v_u~~,~~~~~~~ m_e=
\left[
\matrix{ 0&\delta&\delta'\cr
\delta&0&-\epsilon\cr
\delta'&\sigma+\epsilon&1}
\right]v_d~~.
\label{mquark2}
\eeq  They provide a good fit of the available data in the quarks and the charged lepton sector in
terms of 5  parameters (one of which is complex). In the neutrino sector one obtains a large
$\theta_{23}$ mixing angle,
$\sin^2 2\theta_{12}\sim 6.6\cdot 10^{-3}$ eV$^2$ and $\theta_{13}$ of the same order of
$\theta_{12}$. Mass squared differences are sensitive to the details of the Majorana mass matrix. 

Looking at models with three light neutrinos only, i.e. no sterile neutrinos, from a more general
point of view, we stress that in the above models the atmospheric neutrino mixing is considered
large, in the sense of being of order one in some zeroth order approximation. In other words it
corresponds to off diagonal matrix elements of the same order of the diagonal ones, although the
mixing is not exactly maximal. The idea that all fermion mixings are small and induced by the
observed smallness of the non diagonal $V_{CKM}$  matrix elements is then abandoned. An alternative
is to argue that perhaps what appears to be large is not that large after all. The typical small
parameter that appears in the mass matrices is $\lambda\sim
\sqrt{m_d/m_s}
\sim
\sqrt{m_{\mu}/m_{\tau}}\sim 0.20-0.25$. This small parameter is not so small that it cannot become
large due to some peculiar accidental enhancement: either a coefficient of order 3, or an exponent
of the mass ratio which is less than $1/2$ (due for example to a suitable charge assignment), or the
addition in phase of an angle from the diagonalization of charged leptons and an angle from neutrino
mixing. One may like this strategy of producing a large mixing by stretching small ones if, for
example, he/she likes symmetric mass matrices, as from left-right symmetry at the GUT scale. In
left-right symmetric models smallness of left mixings implies that also right-handed mixings are
small, so that all mixings tend to be small. Clearly this set of models \cite{str} tend to favour
moderate hierarchies and a single maximal mixing, so that the SA-MSW solution of solar neutrinos is
preferred.

\section{Conclusion}

By now there are rather convincing experimental indications for neutrino oscillations.  If so, then
neutrinos have non zero masses. As a consequence, the phenomenology of neutrino masses and mixings
is brought to the forefront.  This is a very interesting subject in many respects. It is a window on
the physics of GUTs in that the extreme smallness of neutrino masses can only be explained in a
natural way if lepton number is violated.  Then neutrino masses are inversely proportional to the
large scale where lepton number is violated. Also, the pattern of neutrino masses and mixings can
provide new clues on the long standing problem of quark and lepton mass matrices. The actual value
of neutrino  masses is important for cosmology as neutrinos are candidates for hot dark matter:
nearly degenerate neutrinos with a common mass around 1- 2 eV would significantly contribute to the
matter density in the universe. 

While the existence of oscillations  appears to be on a solid ground, many important experimental
ambiguities remain.  For
solar neutrinos  it is not yet clear which of the solutions, MSW-SA, MSW-LA, LOW and VO, is
true, and the possibility also remains of different solutions if  not all of the experimental input
is correct (for example, energy independent solutions are resurrected if the Homestake result is
modified).  Finally a confirmation of the LSND alleged signal is necessary, in order to know
if 3 light neutrinos are sufficient or additional sterile neutrinos must be introduced. 
We argued in favour of models with 3 widely split neutrinos. Reconciling large splittings with
large mixing(s) requires some  natural mechanism to implement a vanishing determinant condition.
This can be obtained in the see-saw mechanism if one light right-handed neutrino is dominant, or a
suitable texture of the Dirac  matrix is imposed by an underlying symmetry. In a GUT context, the
existence of  right-handed neutrinos indicates SO(10) at least as a classification group.  The
symmetry group at $M_{GUT}$ could be either (Susy) SU(5) or SO(10)  or a larger group. We have
presented a class of natural models where large right-handed mixings for quarks are transformed
into large left-handed mixings for  leptons by the approximate transposition relation $m_d=m_e^T$
which is approximately realised in  SU(5) models. We have shown that these models can be naturally
implemented  by simple assignments of $U(1)_F$ horizontal charges. 

In conclusion the fact that some neutrino mixing angles are large, while surprising at the start,
was eventually found to be well be compatible, without any major change, with our picture of quark
and lepton masses within GUTs. In fact, it provides us with new important clues that can become
sharper when the experimental picture will be further clarified.

\section{Acknowledgements}
I am grateful to Milla Baldo-Ceolin for inviting me to this exceptionally interesting Workshop and
for her beautiful organisation and splendid hospitality.

\vfill

\begin{thebibliography}{99}
\bibitem{us1} G. Altarelli and F. Feruglio, Phys. Lett. B439(1998)112, hep-ph/9807353.
\bibitem{us2} G. Altarelli and F. Feruglio, JHEP 11(1998)21, hep-ph/9809596.
\bibitem{us3} G. Altarelli and F. Feruglio, Phys. Lett. B451(1999) 388, hep-ph/9812475.
\bibitem{us4} G. Altarelli and F. Feruglio, Phys. Rep. 320(1999)295, hep-ph/9905536.
\bibitem{us5} G. Altarelli, F. Feruglio and I. Masina, Phys. Lett. B472(2000)382, hep-ph/9907532.
\bibitem{us6} G. Altarelli, F. Feruglio and I. Masina,  JHEP 11(2000)040, hep-ph/0007254.
\bibitem{us7} G. Altarelli and F. Feruglio, hep-ph/0102301.
\bibitem{ss} M. Gell-Mann, P. Ramond and R. Slansky in Supergravity, ed. P. van Nieuwenhuizen and D.
Z. Freedman, North-Holland, Amsterdam, 1979, p.315;\\ T. Yanagida, in Proceedings of the Workshop on
the unified theory and the baryon number in the universe, ed. O. Sawada and A. Sugamoto, KEK report
No. 79-18, Tsukuba, Japan, 1979. See also R. Mohapatra and G. Senjanovic, Phys. Rev. Lett. 44, 912
(1980).
\bibitem{sta} I. Stancu, these Proceedings.
\bibitem{eit} K. Eitel, these Proceedings.
\bibitem{mcg} C. McGrew, these Proceedings.
\bibitem{pul} A. Pullia, these Proceedings.
\bibitem{tak} Y. Takeuchi, these Proceedings.
\bibitem{bel} E. Bellotti, these Proceedings.
\bibitem{4nu} See, for example, M. C. Gonzalez-Garcia and C. Pena-Garay, hep-ph/0011245; \\  G. L.
Fogli, E. Lisi and A. Marrone, Phys. Rev. D63:053008, 2001 (hep-ph/0009299);\\ S. M. Bilenkii, C.
Giunti, W. Grimus and T. Schwetz, Phys. Rev. D60:0073007, 1999 (hep-ph/9903454). 
\bibitem{fog} G. L.Fogli, these Proceedings.
\bibitem{cal} D. O. Caldwell, these Proceedings.
\bibitem{moh} R. N. Mohapatra, these Proceedings.
\bibitem{how} P. Horava and E. Witten, Nuc. Phys. B475(1996)94 (hep-th/9603142);\\ N.
Arkani-Hamed, S. Dimopoulos and G. Dvali, Phys. Lett. B429(1998)263 (hep-ph/9803315);\\I.
Antoniadis, N. Arkani-Hamed, S. Dimopoulos and G. Dvali, Phys. Lett. B436(1998)257
(hep-ph/9804398). 
\bibitem{luk} For an immersion into this subject, see, for example, the recent paper by A. Lukas,
P. Ramond, A. Romanino and G. Ross, hep-ph/0011295 and references therein.
\bibitem{lim} C.D. Hoyle et al, Phys. Rev. Lett. 86(2001)1418 (hep-ph/0011014).
\bibitem{an} See, for example, I. Antoniadis and K. Benakli, hep-ph/0007226.
\bibitem{barr} S. M. Barr and I. Dorsner,  hep-ph/0003058.
\bibitem{GG} F. Vissani, hep-ph/9708483;\\ H. Georgi and S.L. Glashow, hep-ph/9808293.
\bibitem{ello} J. Ellis and S. Lola, hep-ph/9904279;\\ J.A. Casas et al, hep-ph/9904395,
hep-ph/9905381, hep-ph/9906281;\\R. Barbieri, G.G. Ross and A. Strumia, hep-ph/9906470;\\ E. Ma,
hep-ph/9907400;\\K.R.S. Balaji et al, hep-ph/0001310 and hep-ph/0002177.
\bibitem{fri} Examples of degenerate models are described in A. Ioannisian, J. W. F. Valle, Phys.
Lett. B332 (1994) 93, hep-ph/9402333;\\  M. Fukugita, M. Tanimoto, T. Yanagida, Phys. Rev. D57
(1998) 4429, hep-ph/9709388; hep-ph/9903499\\  M. Tanimoto, hep-ph/9807283 and hep-ph/9807517;\\ 
H. Fritzsch, Z. Xing, hep-ph/9808272;\\ R. N. Mohapatra, S. Nussinov, hep-ph/9808301 and
hep-ph/9809415;\\ M. Fukugita, M. Tanimoto, T. Yanagida, hep-ph/9809554;\\ Yue-Liang Wu,
hep-ph/9810491;\\  J. I. Silva-Marcos, hep-ph/9811381;\\  C. Wetterich, hep-ph/9812426;\\S.K. Kang
and C.S. Kim, hep-ph/9811379.
\bibitem{hall} R. Barbieri, L. J. Hall, D. Smith, A. Strumia and N. Weiner, hep/ph 9807235.
\bibitem{king} S. F. King, Phys. Lett. B439 (1998) 350(hep-ph/9806440) and hep-ph/9904210;\\ S.
Davidson and S. F. King, Phys. Lett. B445 (1998) 191(hep-ph/9808333);\\Q. Shafi and Z.
Tavartkiladze, Phys. Lett B451(1999)129 (hep-ph/9901243).
\bibitem{jg} H. Georgi and C. Jarlskog, Phys. Lett. B86(1979) 297.
\bibitem{eg} J. Ellis and M. K. Gaillard,  Phys. Lett. B88(1979) 315. 
\bibitem{fro} C. Froggatt and H. B. Nielsen, Nucl. Phys. B147 (1979) 277.
\bibitem{buch} W. Buchmuller and T. Yanagida, hep-ph/9810308.
\bibitem{abelian} P. Binetruy, S. Lavignac, S. Petcov and P. Ramond, Nucl. Phys. B496 (1997) 3,
hep-ph/9610481;\\ N. Irges, S. Lavignac, P. Ramond, Phys. Rev. D58 (1998) 5003, hep-ph/9802334;\\ Y.
Grossman, Y. Nir, Y. Shadmi, hep-ph/9808355.
\bibitem{skp} Y. Hayato et al, (SuperKamiokande Collab.), Phys. Rev. Lett. 83(1999)1529
(hep-ex/9904020).
\bibitem{mp} A. Masiero et al, Phys. Lett. B 115(1982)380;\\B Grinstein, Nucl. Phys.
B206(1982)387;\\ Z. Berezhiani and Z. Tavartkiladze, Phys. Lett. B409 (1997) 220..
\bibitem{faye} P. Fayet, Phys. Lett. B146(1984)41.
\bibitem{kawa} Y. Kawamura, hep-ph/0012125.
\bibitem{hn} L. Hall and Y. Nomura, hep-ph/0103125.
\bibitem{alb} C. H. Albright and S. M. Barr, Phys. Rev. D58 (1998) 013002, hep-ph/9712488;
hep-ph/9901318;hep-ph/0002155; hep-ph/0003251; \\ C. H. Albright, K. S. Babu and S. M. Barr, Phys.
Rev. Lett. 81 (1998) 1167, hep-ph/9802314.
\bibitem{str} See, for example, S. Lola and G. G. Ross, hep-ph/9902283;\\ K. Babu, J. Pati and F.
Wilczek, hep-ph/9912538.



\end{thebibliography}
\end{document}